\documentstyle[11pt]{article}                       %{bk97.tex 17.12.97}

\def\quality{\textwidth 15.5cm \textheight 22cm}
\quality

\def\mlbscale{1pt} %to change: \renewcommand\mlbscale{1.3pt}

\def\Bfig(#1,#2)#3#4{\begin{figure} \begin{center}
    \setlength{\unitlength}{\mlbscale} \begin{picture}(#1,#2) #3
    \end{picture} \end{center} \caption{#4} \end{figure}}
\def\bpic(#1,#2)#3{\setlength{\unitlength}{\mlbscale} 
    \begin{picture}(#1,#2) #3 \end{picture}} 
\def\bline(#1,#2)(#3,#4)(#5){\put(#1,#2){\line(#3,#4){#5}}}

\makeatletter \newcount\@yy
\def\romb(#1,#2)(#3,#4)(#5)
    {\@yy=#5 \divide\@yy by #3 \multiply\@yy by #4
     \multiput(#1,#2)(#5,-\@yy){2}{\bline(0,0)(#3, #4)(#5)}
     \multiput(#1,#2)(#5, \@yy){2}{\bline(0,0)(#3,-#4)(#5)} } 

\def\rect(#1,#2)(#3,#4)
    {\multiput(#1,#2)(0,#4){2}{\bline(0,0)(1,0)(#3)}
     \multiput(#1,#2)(#3,0){2}{\bline(0,0)(0,1)(#4)}}
\makeatother

\def\beq#1#2{\begin{equation} \label{#1} #2 \end{equation}}
\def\bea#1{\begin{eqnarray*} #1 \end{eqnarray*}} \def\a{\!\!\!&\!\!\!\!&}
\def\be{\begin{equation}}   \def\ee{\end{equation}}

\def\function#1{\left\{\!\!\!\begin{array}{ll} #1 \end{array} \right.}
\def\norm#1#2{\left|\left|#1\right|\right|_{#2}}      % ||#1||_{#2}
\def\ns#1{\norm{#1}{\infty}}    \def\nl#1{\norm{#1}{1}}
\def\nv#1{\norm{#1}{_{\rm B\!V}}}

\def\mod1{\,({\rm mod\ } 1)\,}   \def\?#1{}  

\def\BV{{\bf BV\/}}    
\def\la{\lambda}    
   
 \def\Z{{\cal Z}} 
\def\TP{{\rm TP}} \def\PTP{{\rm PTP}} %(Per) turning points

\def\I|#1{{\big|}_{#1}} \def\1#1{\hbox{\large \bf 1}_{#1}} 
\def\lr#1{\left\{ #1 \right\}}  \def\lrp#1{\left( #1 \right)}

\def\n{\noindent}  \def\ep{\varepsilon}  \def\var{{\rm var}}
\def\IP{{\bf P}}         
\def\IZ{\hbox{{\rm Z}\kern-.3em{\rm Z}}}
\def\IR{\hbox{\rm I\kern-.2em\hbox{\rm R}}}
\def\IN{\hbox{I\kern-.2em\hbox{N}}}
\def\proof{\smallskip \noindent {\bf Proof. \ }}
\newcommand\filledsquare{\ \vrule width 1.5ex height 1.2ex}  %filled square
\def\qed{\hfill\filledsquare\linebreak\smallskip\par}

\newtheorem{theorem}{Theorem}[section]   %Numbering: Theorem--Other section
\newtheorem{lemma}{Lemma}[section]       %{lemma}[theorem]{Lemma}   section
\newtheorem{proposition}[lemma]{Proposition} %lemma
\newtheorem{definition}{Definition}[section] %{lemma}[theorem]{Lemma} section
             \catcode`@=11 \@addtoreset{equation}{section} \catcode`@=12

\begin{document} 
%%%%%%%%%%%%%%%%%%%%%%%%%%%%%%%%%%%%%%%%%%%%%%%%%%%%%%%%%%%%%%%%%

\title{Random perturbations of chaotic dynamical systems. \\
       Stability of the spectrum}
\author{Michael Blank\thanks{
         On leave from Russian Academy of Sciences, Inst. for 
         Information Transmission Problems, 
         B. Karetnij Per. 19, 101447, Moscow, Russia, blank@obs-nice.fr}
        , Gerhard Keller
        \\ \\ 
         Mathematisches Institut, Universitat Erlangen-Nurnberg \\
         Bismarckstrasse 1 1/2, D-91054 Erlangen, Germany.
        }
\date{December, 1997}
\maketitle
\n {\bf Abstract} -- For piecewise expanding one-dimensional maps 
without periodic turning points we prove that isolated eigenvalues of small 
(random) perturbations of these maps are close to isolated eigenvalues of 
the unperturbed system. (Here ``eigenvalue'' means eigenvalue of the 
corresponding Perron-Frobenius operator acting on the space of functions of 
bounded variation.) This result applies e.g.\ to the approximation of the 
system by a finite state Markov chain and generalizes Ulam's conjecture 
about the approximation of the SBR invariant measure of such a map. We 
provide several simple examples showing that for maps with periodic turning 
points and for general multidimensional smooth hyperbolic maps isolated
eigenvalues are typically unstable under random perturbations. Our main 
tool in the 1D case is a special technique  for ``interchanging'' 
the map and the perturbation, developed in our previous paper~\cite{BK95}, 
combined with a compactness argument. 
\bigskip

%%%%%%%%%%%%
\section{Introduction} \label{Int}

We discuss stochastic stability in the following general framework:
A discrete time dynamical system is a pair $(f,X)$, where 
$X \subset \IR^d$ is a bounded phase space (say $X=[0,1]^d$) and $f:X \to X$ 
is a nonsingular map, iterations of which define trajectories of the system. 
Nonsingular means that $m(f^{-1}A)>0$ for any measurable set $A \subseteq X$ 
with positive Lebesgue measure $m(A)>0$.

Consider now small {\em random perturbations} of the discrete time dynamical 
system. Roughly speaking this means that, when we apply $f$ to a point 
$x \in X$, rather than choosing the exact value of $fx$ we choose in a random 
way, in accordance with some distribution, a point from the ball
$B_\ep(fx)$ (i.e. with centre at the point $fx$ and radius $\ep$). 

\begin{definition} Let $Q_\ep(x,A)$ be a family of transition probabilities
and $f:X\to X$ a map. We denote by $f_\ep$ the Markov process on the phase 
space $X$ defined by the transition probabilities $Q_\ep(fx,A)$ and call 
$f_\ep$ a {\em random perturbation} of $f$. \end{definition}

Our main assumptions on the perturbations are the following:
\beq{local-cond}{Q_\ep(x,A) = 0 \quad\hbox{if   dist}(x,A) > \ep 
                                \qquad\quad\qquad\quad({\rm locality}),}
\beq{small-ass}{ \nl{Q_\ep h - h} \le {\bf d}(Q_\ep) \cdot \var(h)\to0
                 \hbox{  as  } \ep\to0  \quad({\rm smallness}),}
\beq{var-as}{ \var(Q_\ep h) \le \var(h) + C \nl{h} 
                                \qquad\qquad\qquad\quad ({\rm regularity})}
for any function $h$ of bounded variation ($\var(h) < \infty$). The 
parameter $\ep$ here plays the role of a ``magnitude'' of the perturbation.
These assumptions are satisfied for a broad class of random perturbations, 
among them convolutions with absolutely continuous transition probabilities,
bistochastic absolutely continuous perturbations, singular perturbations of 
point mass type, deterministic perturbations by chaotic maps close to 
identity, and Ulam type perturbations (see, for example, \cite{BK95}
for details).

Stochastic stability of the Sinai-Bowen-Ruelle (SBR) measure of a
dynamical system discussed under the same assumptions in \cite{BK95}
may be considered as a weak kind of stability, because other 
statistical characteristics may be unstable. In this paper we further 
explore the problem of stochastic stability and study the stability of
the spectrum of the Perron-Frobenius operator $\IP_f$ (PF-spectrum for
brevity), considered as an operator on the Banach space $(\BV,\nv{\cdot})$ 
of functions of bounded variation. This operator describes the
dynamics of densities under the action of the map $f$. Our main
stability result provides rather general sufficient conditions for 
stochastic stability of isolated eigenvalues of the spectrum in the 
one-dimensional piecewise expanding case, and counterexamples below
show that these conditions can hardly be relaxed. Moreover we
demonstrate that arbitrarily small random perturbations and especially 
Ulam type perturbations (see below) of a generic multidimensional 
hyperbolic map (having a stable manifold) can completely change the 
PF-spectrum of the system.

One of the main motivations for the present work was to study the stability 
of the PF-spectrum in Ulam's construction of a finite state Markov chain 
approximation \cite{Ul,BK95} of chaotic dynamics. The idea of the construction 
is to take a finite partition $\{\Delta_i\}$ of the phase space with bounded 
volume ratios and to approximate the action of the map $f$ by the Markov 
chain with transition probabilities
$$ p_{ij}:=\frac{|\Delta_i \cap f^{-1}\Delta_j|}{|\Delta_i|}.$$
This construction can be considered as a special type of small random 
perturbations, where the transition operator satisfies all above assumptions. 
The convergence of invariant measures of these finite Markov chains to
the SBR measure of the approximating dynamical system under this
construction was proved in \cite{Li} for piecewise expanding (PE) maps
with large enough expanding constants, and for general PE maps in 
\cite{BK95,Bl20} (see also a review there). For numerical applications
of this method see e.g. \cite{dellnitz}. Since Ulam's construction 
provides a rather general approach for numerical modeling of chaotic 
dynamics, the question of stability of the PF-spectrum and the study
of nonconvolution type random perturbations becomes important not only
from a purely theoretical point of view but also from a practical one. 

The operator $\IP_f$ preserves integrals, thus its leading eigenvalue is 
equal to 1 and the most important feature of its spectrum $\Sigma(\IP_f)$ 
is the modulus of the second largest spectral value 
$r_2:=\sup\{ |\tau|: \; \tau \in \Sigma(\IP_f), \; |\tau|<1\}$, which
characterizes the rate of convergence to the SBR measure. The
corresponding value for the perturbed operator is denoted by $\tilde r_2$.  

The basic idea here is the following. Recall that $\nv{h}=\var(h)+\nl{h}$. 
In \cite{BK95} we proved that the transition operator $P_\ep = Q_\ep \IP_f$ 
of the randomly perturbed map satisfies the uniform Lasota-Yorke type 
inequality
\beq{LY-type}{ \nv{P_\ep^N h} \le \alpha\cdot\nv{h} + C\cdot\nl{h} }
for $h \in \BV$, some fixed integer $N$, $\alpha\in(0,1)$ and $C>0$ 
independent of $\ep$. This yields at once the existence of $f_\ep$-invariant 
densities $h_\ep$ with $\nv{h_\ep} \le \frac C{1-\alpha}$, such that 
(\ref{small-ass}) forces each weak limit $h_*=\lim_{\ep \to 0}h_\ep$ to be 
an invariant density for $f$. (In fact, as $\nv{h_\ep}$ is uniformly bounded, 
$h_*$ is not only a weak limit, but $\nl{h_*-h_\ep} \to 0$ as $\ep \to 0$.) 
If the expanding constant $\la$ is larger than $2$, (\ref{LY-type}) was 
proved with $N=1$ in various settings, see e.g. \cite{{Ke1},{Bl5},{Ki3}} 
and references therein. For quite a while it was supposed that the extension 
of this inequality to cases with $\la \in (1,2]$ is only a technical problem. 
However, the counterexample constructed in \cite{Bl17} shows that the 
situation is not so simple. After this counterexample it became clear, that 
the main problem is the possible existence of periodic {\em turning} points, 
i.e. points where the derivative of the map $f$ is not well defined. 
Namely, under the action of random perturbations ``traps'' or ``absorbing 
sets'' can appear near these periodic turning points, which leads to the
appearance of new localized ergodic components in the perturbed system. 

\begin{definition} Let $X=[0,1]$. A map $f:X\to X$ is {\em piecewise $C^2$}
if there exists a partition of $X$ into disjoint intervals $\{X_j\}$, such 
that $f\I|{{\rm Clos}(X_j)}$ is a $C^2$-diffeomorphism (of the closed 
interval ${\rm Clos}(X_j)$ to its image). Its {\em expanding constant} is 
defined as
$$ \la_f := \inf_{j, \, x \in X_j} |f'(x)|. $$
A piecewise $C^2$ map is called {\em piecewise expanding} (PE), if 
$\la_{f^k}>1$ for some iterate $f^k$.
\end{definition}

\begin{definition} The {\em image} of a measure $\mu$ under the action of a
map $f$ is the measure $f\mu$ defined by $f\mu(A)=\mu(f^{-1}A)$ for any
measurable set $A$. By $f_\ep \mu$ we mean the measure $f_\ep \mu(A)=\int
Q_\ep(fx,A)d\mu(x)$. A measure $\mu$ is $f$\ ($f_\ep$)--invariant, if
$f\mu=\mu$\ ($f_\ep\mu=\mu$). $\mu$ is called {\em smooth}, if it has a
density with respect to Lebesgue measure. \end{definition}

\begin{definition} A {\em turning point} of a map $f$ is a point, where the 
derivative of the map is not well defined. We denote the set of turning 
points by $\TP$, and the set of periodic turning points by $\PTP$. 
\end{definition}

Recall that the {\em essential spectral radius} of the operator
$\IP_f$ is the smallest nonnegative number $\theta$ for which elements
of the spectrum $\Sigma(\IP_f)$ outside of the disk of radius $\theta$
centered at the origin are isolated eigenvalues of finite
multiplicity. In \cite{Ke2} it was shown that for a PE map 
$\theta=\lim_{n \to \infty} \root n \of {\la_{f^n}^{-1}}$. 
Therefore $f$ is PE if and only if $\theta<1$. 
From now on we fix some numbers $\theta',\tilde\theta$ arbitrarily
close to $\theta$ with $\theta<\theta'<\tilde\theta<1$.

\begin{theorem} \label{is-eig-stab} Let $f$ be a piecewise expanding map 
with $\PTP=\emptyset$, and let perturbations $Q_\ep$ satisfy 
(\ref{local-cond}), (\ref{small-ass}) and (\ref{var-as}). Let $r$ be
an accumulation point of eigenvalues $r_\ep>\tilde\theta>\theta$ of
the perturbed operators $P_\ep:=Q_\ep\IP_f$ for $\ep \to 0$. Then there
are a sequence $k \to \infty$ and a function $h \in BV$ such that 
$r_{\ep_k} \to r$ and $\nl{h_{\ep_k} - h}\to 0$ as $k\to\infty$, where
$\IP_f h =r h$ and $P_{\ep_k}h_{\ep_k}=r_{\ep_k}h_{\ep_k}$. \end{theorem}

In other words, any accumulation point of the eigenvalues of the perturbed 
operators lying outside of the disk containing the essential spectrum of 
the original operator is an isolated eigenvalue of $\IP_f$. This generalizes 
the corresponding part of the results for convolution type perturbations of 
smooth expanding maps in \cite{BY}.
In the case of Ulam type approximations based on Markov partitions of smooth
hyperbolic maps our theorem is complemented by the following result 
\cite{Fr2}: Each isolated eigenvalue is a limit point of eigenvalues 
of the corresponding Ulam operators.
 
It is worth to remark that the rate of convergence of the eigenfunctions 
for the eigenvalue $1$ is $O(\ep|\log\ep|)$ whereas
we have no rates of convergence for other spectral quantities.
In particular, in some cases
numerical experiments show very slow convergence of the
$r_\ep$ to true eigenvalues.

Technically the proof of this theorem is based on the following proposition, 
being a consequence of several technical results obtained in \cite{BK95}. 

\begin{proposition} \label{LY-noise} Let $f$ be a piecewise expanding map 
with $\PTP=\emptyset$, and let perturbations $Q_\ep$ satisfy 
(\ref{local-cond}), (\ref{small-ass}) and (\ref{var-as}). Then there 
exist constants $C, \ep_0$ such that for some finite $N$
$$ \var(P_\ep^N h) \le \tilde\theta^N \var(h) + C \nl{h} $$
for any $\ep \in (0,\ep_0)$ and any function $h \in \BV$. \end{proposition}

The paper is organized as follows. In Section~2 we recall some
necessary definitions and prove our main stability result. Section~3
is devoted to the analysis of various situations when the spectrum is
not stable with respect to random perturbations. Especially important
among these situations are the spectrum collapse in the absence of
isolated eigenvalues and the instability of the spectrum of
multidimensional hyperbolic maps due to the presence of the stable
foliation of the map. Finally we discuss a possible generalization of
the notion of the spectrum by means of zero-noise limit of the spectra
of randomly perturbed systems.

%%%%%%%%%%%%
\section{Proof of the stability result}

Recall that the variation of a function over $h:X=[0,1] \to \IR_1$ is
defined as $\var(h) := \sup\lr{\int_X \phi'h\,dx}$, where the supremum
is taken over all functions $\phi \in C^1(\IR)$ with compact support, 
$\ns{\phi} \le 1$ and $\ns{\phi'}<\infty$. Notice that as $X=[0,1]$ 
is bounded, $\nl{h} \le \frac12 \var(h)$ for all $h\in\BV$, and if 
$I\subseteq X$ is an interval, then $\var(h\cdot \1I)\le\var(h)$. Indeed, our 
setting means that $\var(h)$ is the variation of $h \cdot \1X$ over $\IR$.

Given $N$ and $\beta>0$ as defined above we refine the partition $\Z$ 
into intervals of monotonicity of the map $f$ by 
adding further points to $\TP$ in such a way that 
\beq{beta-ass}{
   \var\left(\la_{f|Z}^{-1}f_{|Z}'\right),\ 
   \var\left((\la_{f|Z} f_{|Z}')^{-1} \right) \le \beta \ ,}
not introducing new $\TP$ that are mapped to other $\TP$.

Define $\tilde P_1,\tilde P_2:\BV\to \BV$ by
$$ \tilde P_1 h=Q_\ep \IP_f(h\cdot \1{X\setminus Y}),\quad
   \tilde P_2 h=Q_\ep \IP_f(h\cdot \1Y)$$
where $Y$ is a neighbourhood of \TP \ scaling linearly with respect to $\ep$
(see \cite{BK95}).

\begin{proposition}\label{P1-P2-prop}\cite[Proposition 3.1]{BK95}
Suppose that there are constants $C_1,C_2>0$ and $\alpha\in(0,1)$ such 
that
\beq{1D-LY-decomp}{
   \var(\tilde P_j^kh)\le C_1\alpha^k\var(h)+C_2 \nl{h} \quad 
        \mbox{ for all } k \in \IZ_+ }
and that there is some $N \in \IZ_+$ such that 
\beq{1D-transitions}{
   \tilde P_2\tilde P_1^k\tilde P_2=0 \quad\mbox{ for all $k=1,\ldots,N$.} }
Then
\bea{ \var((Q_\ep \IP)^N h) \le
   \a \left(\frac{N(N+1)}{2}C_1^3+C_1\right)\alpha^N\cdot\var(h) \\
   \a + (1+C_1+C_1^2)C_2\frac{(N+1)^2}2\cdot \nl{h} \ .}
\end{proposition}

The assumptions (\ref{1D-LY-decomp}) and (\ref{1D-transitions}) were 
verified in \cite{BK95} for PE maps with $\PTP=\emptyset$ and for 
sufficiently small $\ep$ without paying attention to a particularly
sharp estimate of the constant $\alpha$. For the purposes of our
present paper we need to show that $\alpha$ can be chosen as $\alpha=\theta'$,
i.e.\  close to the essential spectral radius $\theta$.

Since $\PTP=\emptyset$, there is some $n_0$ such that $\tilde P_2^{n_0}=0$
so that inequality (\ref{1D-LY-decomp}) can be satisfied for any
positive $\alpha$ with suitable constants $C_1,C_2$ depending only 
on $f$ and $Q$. For $\tilde P_1$ the relevant estimate is given in 
\cite[Proposition 3.8]{BK95}. The value $\alpha=(\frac 34)^{1/N}$
given in the statement of the proposition is not as sharp as the 
corresponding proof permits. In fact, the effictive estimate derived
in the proof is
\bea{ \a \var(\tilde P_1^N h) \\
   \a\le (1 + \frac 32 \beta)^{2N} (\lambda_{f^N}^{-1} 
       + \frac 12 N\beta\la_f^{-N}) \var(h)
 + (\lambda_{f^N}^{-1} + \frac 12 N\beta\la_f^{-N}) \tilde C_N \cdot \nl{h} \ }
for each $h\in\BV$.
\footnote{Observe that the brackets in the corresponding formula in 
\cite{BK95} are set slightly wrong, which had no further effect on the 
subsequent proofs in that paper.}

\bigskip

\n {\bf Proof} of Proposition~\ref{LY-noise}. It remains to choose the constant 
$\beta>0$ so small that
$$ (1 + \frac 32 \beta)^{2N} (\theta^N 
       + \frac 12 N\beta\la_f^{-N}) < (\theta')^N < \tilde\theta^N $$
to obtain the statement of Proposition~\ref{LY-noise}. \qed

\bigskip

\n {\bf Proof} of Theorem~\ref{is-eig-stab}.  
Suppose that $P_\ep h_\ep = r_\ep h_\ep$ for some $h_\ep \in BV$ 
with $\nl{h_\ep}=1$ and $|r_\ep|\ge\theta'$. Then there is a constant 
$S>0$ depending only on $\tilde\theta$ and on the constants in the 
Lasota-Yorke type inequality such that $\var(h_\ep)\le S$. Indeed,
by Proposition~\ref{LY-noise}
$$ \var(P_\ep^N h_\ep) \le \tilde\theta^N \var(h_\ep) + C \nl{h_\ep} .$$
Since $P_\ep^N h_\ep = r_\ep^N h_\ep$, we obtain the following estimate:
$$ \var(h_\ep) = \var(\frac1{r_\ep^N}P_\ep^N h_\ep) 
   \le \left( \frac{\tilde\theta}{r_\ep}\right)^N \var(h_\ep) 
 + \frac{C}{r_\ep^N} \nl{h_\ep} ,$$
or
$$ \var(h_\ep) \le \frac{C}{r_\ep^N - \tilde\theta^N} := S < \infty, $$
since $r_\ep > \tilde\theta$. In particular also 
$\var(\IP_f h_\ep)\le S$ and $\var(\IP_f h)\le S$, where $h=h_0$.

Let $r$ be any accumulation point of the $r_\ep$ for $\ep \to 0$. Then 
there are a sequence $\ep_k \to 0$ and a function $h \in BV$ such that 
$r_{\ep_k} \to r$ and $\nl{h_{\ep_k} - h}\to 0$ as $k\to\infty$. In 
particular $\var(h)\le S$. Using the assumption~(\ref{small-ass}) it 
follows that
$$  \nl{\IP_f h - P_{\ep_k} h_{\ep_k}}
    \le \nl{(\IP_f - P_{\ep_k})h} + \nl{P_{\ep_k}(h-h_{\ep_k})} $$
$$  \le {\bf d}(Q_{\ep_k}) \cdot S + \nl{h-h_{\ep_k}} \to 0 $$
as $\ep \to 0$ and
$$  \nl{r h - r_{\ep_k} h_{\ep_k}}
    \le |r| \cdot \nl{h-h_{\ep_k}} + |r-r_{\ep_k}|\cdot\nl{h_{\ep_k}} \to 0\ .$$
Hence $\IP_f h = r h$.  \qed

\?{ This means that eigenvalues and eigenfunctions of the perturbed
  system converge to those of the unperturbed one. With a bit more
  care one should be able to prove that the multiplicity of the
  limiting eigenvalue for $P$ is at least as big as the
  $\limsup_{r_{\ep_i} \to r}$ of the multiplicities of the
  approximating eigenvalues of the $P_\ep$. What seems more difficult
  to prove is that also the eigenprojections converge in some sense
  and that each eigenfunction of the unperturbed system is
  approximated by eigenfunctions of the perturbed system.}

%%%%%%%%%%%%%%%%%%%%%%%%%%%%%%%
\section{Instability of the PF-spectrum for generic maps} 

\subsection{Instability of the essential spectrum}

The above estimates ensure stochastic stability of the isolated
eigenvalues of $\IP_f$ in the sense that spectral points outside the essential
spectrum cannot arise ``from nothing'' under perturbations.
We have no examples of maps $f$ without \PTP \ and perturbations $Q_\ep$
where an isolated eigenvalue of $\IP_f$ is not approximated by
eigenvalues of the $P_\ep$. However, if the
only isolated eigenvalue is $1$, the essential spectrum may
collapse, which leads to spectrum localization ($\tilde r_2=0$). 
The simplest example when this phenomenon takes place is the
case of Ulam perturbations of the well known dyadic map $x \to
2x \mod1$, when the number of intervals in the Ulam partition is
equal to $2^n$. If this number is not an integer power of $2$
there will be a stable eigenvalue $\tilde r_2=1/2$ (on the
boundary of the essential spectrum) with the eigenvector ($x-1/2$).
The following statement generalizes this observation.

\begin{theorem}
  Let $f:X\to X\subset\IR^d$ be piecewise linear with
  respect to a partition $\Z$ and assume that $f(I)=X$ for each
  $I\in\Z$.
  Denote by $\Z_N$ the partition of $X$ into domains of linearity of $f^N$. 
  Then the spectrum of the Ulam
  operator $P_{\Z_N}$ constructed with respect to the partition $\Z_N$
  consists of only $0$ and $1$.
\end{theorem}

Sketch of {\bf proof}: 
Let $h$ be a piecewise constant function on $\Z_N$ such that $\int
h(x)\,dx=0$. Then $P_{\Z_N}^Nh=0$, which proves that $\tilde r_2=0$.

\subsection{Instability in the periodic turning points case}

In this more general case we cannot proceed as above in spite of the
fact that the first eigenvector is stable under some additional 
restrictions (see \cite{BK95} for details). The reason is that the proof of
this statement in \cite{BK95} uses slightly weaker estimates of the
rate of convergence, which do not ensure our Proposition~\ref{LY-noise}. 
In this section we want to argue that isolated eigenvalues of the
spectrum (except 1) might be unstable indeed. To show this we modify the
W-map (described in detail in \cite{BK95}) such that locally around
the fixed turning point $c$ it remains the same, while no point
``outside'' is mapped into a small neighborhood of the point $c$. The 
modified W-map is shown in the Figure~\ref{M-W-map}. Observe that
this map is Markov. Consider the following random perturbation: 
$x \to x+\ep$ with probability $1-\delta$ and $x \to x$ with
probability $\delta$. We apply this random perturbation only in the 
neighborhood of the point $c$ consisting of two neighbouring intervals
of monotonicity. For $\delta=0$ there is a $\ep$-small ergodic
component $I_\ep$ around the point $c$. For small enough
$\ep,\delta>0$ the escape rate out of the set $I_\ep$ is of 
order $\delta$. By the construction of the map, the probability to hit into 
the interval $I_\ep$ starting outside of it is equal to zero. Therefore 
the transition operator corresponding to the randomly perturbed system 
must have an eigenvalue of order $1-\delta$. Thus the spectral gap of
the perturbed operator vanishes as $\delta \to 0$.

\bigskip
{\bf Example} 1. Let us discuss now in more detail Ulam
perturbations of the modified W-map. Let $x$ be a local
coordinate such that the fixed turning point be $0$, and denote the Ulam
interval containing this point by $I=[-a,b]$. Define $\la=|f'_{|_I}|$. 
Then the probability to remain in
the inteval $I=[-a,b]$ in Ulam's approximation is equal to 
$$ p(a,b) = \frac{a/\la + \min\lr{a, \la b}/\la}{a+b} 
          = \frac{a + \min\lr{a, \la b}}{\la(a+b)}.$$
Let $a=\la b$. Then 
$$ p(\la b, b) = \frac{\la b + \la b}{\la(\la b + b)} 
               = \frac2{\la+1} = 1 - \frac{\la-1}{\la+1} .$$
On the other hand, if $a \le b$ then the image of any other Ulam interval 
does not intersect with the interval $I$. This shows that
$2/(\la+1)<1$ belongs to the spectrum of the perturbed operator,
whereas the special form of the modified W-map implies that the
spectrum of $\IP_f$, except for the eigenvalue $1$,
is contained in the circle with radius $1/\la<2/(\la+1)$. Observe that 
in this case the perturbation cannot make the spectral gap 
arbitrarily small, but $r_2$ might be unstable.
In any case this proves the appearance of an eigenvalue of the perturbed 
operator in the ``spectral gap'' of the unperturbed one.

%%%%%%%%%%%%%%%%%%%%%%%%%%%%%%%%%%%%%%%%%%%%%%%%%%%%%%%%%%%%%%%%%%
%% Picture for the modified W-map and 2D Ulam's partitions.
\Bfig(150,150)
      {\bline(0,0)(1,0)(150)   \bline(0,0)(0,1)(150)
       \bline(0,150)(1,0)(150) \bline(150,0)(0,1)(150)
       \bline(0,0)(1,1)(150)
       \bline(37.5,0)(0,1)(37.5) \bline(30,0)(0,-1)(5) \bline(42,0)(0,-1)(5) 
       \bline(0,75)(1,0)(150)  \bline(75,0)(0,1)(150) \thicklines
       \bline(37.5,37.5)(-2,-3)(25)\bline(37.5,37.5)(2,-3)(25)
       \bline(0,150)(1,-6)(12.5) \bline(75,150)(-1,-6)(12.5)
       \bline(112.5,75)(-1,2)(37.5) \bline(112.5,75)(1,2)(37.5)
       \put(35,39){$c$} \put(35,-9){$I$}}
{The modified W-map. \label{M-W-map}}
%%%%%%%%%%%%%%%%%%%%%%%%%%%%%%%%%%%%%%%%%%%%%%%%%%%%%%%%%%%%%%%%%%

\bigskip
One might argue that still it is possible that under the action of
well behaving random perturbations of convolution type the influence of 
just a few periodic turning points may not matter. To show that this is 
not the case, consider the following symmetric random perturbation: 
$$ x \to \function{x-\ep &\mbox{with probability } q \\  
                   x     &\mbox{with probability } 1-2q \\
                   x+\ep &\mbox{with probability } q ,} $$
with $0<q \ll 1$. We apply this pertutbation to the same modified W-map for 
$\la=2$ (notice, that in this case $r_2=1/2$), and show that the 
corresponding transition oprator $Q_\ep \IP$ has 
an eigenvalue $1-2q > 1/\la$. Since no point outside of a small neighborhood 
of the fixed point can hit into this neighborhood, it is enough (as in our 
previous examples) to study the escape rate from this neighborhood. If 
$\la=2$ then locally the behaviour of the randomly perturbed system is 
completely described by the following random walk model on $\IZ$:
$$ x \to \function{ 2x + \xi  &\mbox{if } x \ge 0 \\
                   -2x + \xi  &\mbox{otherwise} ,} $$
$$ \xi = \function{-1 &\mbox{with probability } q \\  
                    0     &\mbox{with probability } 1-2q \\
                    1 &\mbox{with probability } q .} $$
Clearly, if $|x|>1$ then the trajectory of this point will never return 
to zero. Consider the part of the transition matrix corresponding to the 
points $-1$, $0$ and $1$. It can be written as
$$ \lrp{ \begin{array}{ccc} q &0 &0 \\ q &1-2q &q \\ q &0 &0 \end{array}} $$
Thus this matrix has an eigenvalue $1-2q$, which proves our statement.

\subsection{Instability for a generic multidimensional hyperbolic map}

Stability of spectral properties becomes a much more delicate problem 
in the multidimensional case. Traditionally, to define this spectrum one 
considers the Perron-Frobenius operator for the expanding map defined on  
unstable manifolds induced by the original map (see, for example, 
\cite{Yo,Fr2,BTV}). Another way to calculate the isolated eigenvalues
is to study the so called weighted dynamical $\zeta$-function of a
map, which counts periodic points of the map weighted by Jacobians in
the unstable direction. Zeros and poles of the $\zeta$-function correspond to
isolated eigenvalues of the map (see, for example, \cite{Ba} and
references therein). 

Our spectral stability results can be generalized for finite
systems of weakly coupled 1D PE maps (see \cite{Bl20} for definitions). 
On the other hand, even for a more general multidimensional PE map our
construction does not work, since there is no good control over
coefficients of a Lasota-Yorke type inequality in this case, contrary 
to the 1D case.

\bigskip
{\bf Example} 2.  Consider a smooth hyperbolic map $f: \IR^2 \to \IR^2$. 
Let $0$ be a hyperbolic fixed point of the map, and let the horizontal
direction be locally unstable with the expanding constant $\la_u>1$,
while the vertical direction be contracting with $\la_s \ll 1$. We
consider Ulam partitions into equal squares rotated by the 
angle $\pi/2$ with respect to the coordinate axes. One element of the
partition together with its preimage is shown in Figure~\ref{rot-Ulam}. 
Straightforward calculations show that the probability to remain in the 
considered Ulam square is of order 
$$ p {\buildrel {(\la_s\ll1)} \over \approx} 1 - \lrp{1 - \frac1{\la_u}}^2 
     = \frac1{\la_u} \lrp{2 - \frac1{\la_u}} = \frac34 \big|_{\la_u=2} .$$
One can calculate this probability exactly also for the case of finite
values of $\la_s$, for example $p=2/3$ for $\la_s=1/2$.
This is not sufficient to prove that $\tilde r_2 > r_2$, but it indicates that 
the limit behaviour of the approximation might differ from that of the 
original map. A slightly less striking example of this type was discussed 
in \cite{Ki4}, where it was claimed that the worst situation is when the 
angle between the contracting and expanding directions is small.

%%%%%%%%%%%%%%%%%%%%%%%%%%%%%%%%%%%%%%%%%%%%%%%%%%%%%%%%%%%%%%%%%%
%% Picture for the 2D rotated Ulam's partition.
\Bfig(150,150)
      {\bline(0,0)(1,0)(150)   \bline(0,0)(0,1)(150)
       \bline(0,150)(1,0)(150) \bline(150,0)(0,1)(150)
       \bline(0,75)(1,0)(150)  \bline(75,0)(0,1)(150)
       \put(135,75){\vector(1,0){6}}
       \put(15,75){\vector(-1,0){6}}
       \put(75,20){\vector(0,1){6}} \put(75,130){\vector(0,-1){6}}
       \thicklines
       \bline(40,75)(1,1)(35)  \bline(40,75)(1,-1)(35)
       \bline(110,75)(-1,1)(35)  \bline(110,75)(-1,-1)(35)
       \bline(50,75)(1,3)(25)  \bline(50,75)(1,-3)(25)
       \bline(100,75)(-1,3)(25)  \bline(100,75)(-1,-3)(25)
       \put(130,79){$\la_u$}}
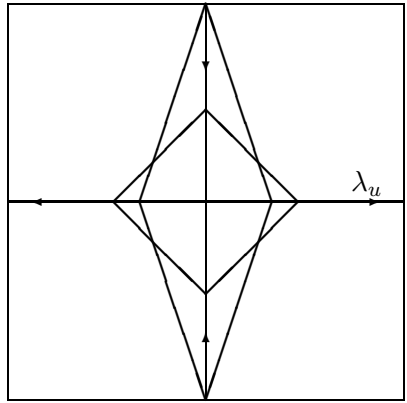
{One element of the rotated Ulam partition and its preimage in 
the 2D hyperbolic case. \label{rot-Ulam}}
%%%%%%%%%%%%%%%%%%%%%%%%%%%%%%%%%%%%%%%%%%%%%%%%%%%%%%%%%%%%%%%%%

\bigskip

The following numerical example shows that the untypical 
instability of the essential spectrum due to the 
presence of periodic turning points in the one-dimensional case 
becomes typical for multidimensional maps. Near a periodic point of a 
multidimensional hyperbolic map stable and unstable foliations are 
coming arbitrarily close one to another. Therefore an arbitrarily small 
(random) perturbation can mix them (similarly to the situation near 
periodic turning points (see \cite{BK95})). We study an example as simple
as possible to demonstrate that this type of behavior is generic. 
\?{Notice that in  examples Ulam's conjecture about the convergence of
invariant measures of the approximating finite Markov chains to the
SBR measure of the original map holds.}

\bigskip 
{\bf Example} 3.  Consider the well known ``cat'' map, which is the simplest 
example of a smooth two dimensional hyperbolic map. This is a map from 
the unit torus $X=[0,1] \times [0,1]$ into itself defined by  
$(x,y) \mapsto (x+y \mod1, x+2y \mod1)$. We consider two partitions of $X$ 
into equal squares. First we simply divide horizontal and vertical axes 
into $n$ equal intevals, whose products give a partition into $N=n^2$ 
squares, which we call the standard partition. There is a one to one 
correspondence of these squares and pairs of integers $(i=nx, j=ny)$, where 
$(x,y)$ is the pair of coordinates of the lower left corner of a 
square. Here $i,j \in \{0,1,\dots,n-1\}$. Simple calculation gives the 
following transition probabilities for the corresponding Markov chain 
whose elements are numbered as $jn+i+1$ (see also Figure~\ref{Im-Stand-part}):

\begin{center}
\begin{tabular}{|c|c|c|c|c|} \hline
  {}   & $i+j,i+2j$& $i+j,i+2j+1$& $i+j+1,i+2j+1$& $i+j+1,i+2j+2$ \\ \hline
 $i,j$ & $1/4$     & $1/4$       & $1/4$         & $1/4$    \\ \hline
\end{tabular} \end{center}

%%%%%%%%%%%%%%%%%%%%%%%%%%%%%%%%%%%%%%%%%%%%%%%%%%%%%%%%%%%%%%%%%%
%% Picture for the ``rectangular'' partition.
\Bfig(160,110)
      {\rect(0,0)(160,110) \rect(10,40)(30,30)
       \rect(90,10)(30,90) \rect(120,10)(30,90)
       \bline(90,40)(1,0)(60) \bline(90,70)(1,0)(60)
       \thicklines \put(50,55){\vector(1,0){20}}
       \bline(90,10)(1,1)(30)    \bline(120,40)(1,2)(30)
       \bline(150,100)(-1,-1)(30) \bline(90,10)(1,2)(30)
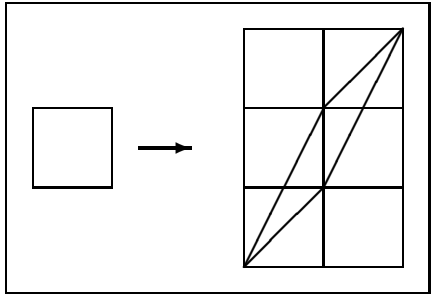
      }{Image of an element of the ``standard'' partition by the
        ``cat'' map. \label{Im-Stand-part}}
%%%%%%%%%%%%%%%%%%%%%%%%%%%%%%%%%%%%%%%%%%%%%%%%%%%%%%%%%%%%%%%%%

Moduli of the ``second'' eigenvalues ($r_2$) of the transition matrices 
($n^2 \times n^2$), and their multiplicities (in parenthesis) are
shown in the following table:

\begin{center} \begin{tabular}{|r|l||r|l||r|l||r|l||r|l|} \hline
$n$& $r_2$& $n$& $r_2$& $n$& $r_2$& $n$& $r_2$& $n$& $r_2$ \\ \hline
 2 &0.0000(3 )& 3 &0.3536(8 )& 4 &0.0000(15)& 5 &0.3299(20)& 6 &0.3536(8 )\\ 
 7 &0.4886(16)& 8 &0.4454(24)& 9 &0.3847(24)& 10&0.4275(12)& 11&0.5161(20)\\
12 &0.3783(48)& 13&0.4835(28)& 14&0.4886(16)& 15&0.4045(16)& 16&0.4454(24)\\
17 &0.4335(24)& 18&0.5957(24)& 19&0.5387(36)& 20&0.4275(12)& 21&0.5357(32)\\
\hline \end{tabular} \end{center}

\?{\begin{center} \begin{tabular}{|r|l||r|l||r|l||r|l||r|l|} \hline
$n$& $r_2$& $n$& $r_2$& $n$& $r_2$& $n$& $r_2$& $n$& $r_2$ \\ \hline
 3 &0.3536(8 )& 4 &0.0017(6 )& 5 &0.3299(20)& 6 &0.3536(8 )& 7 &0.4886(16)\\ 
 9 &0.3847(24)& 10&0.4275(12)& 11&0.5161(20)& 12&0.3783(48)& 13&0.4835(28)\\
 14&0.4886(16)& 15&0.4045(16)& 17&0.4335(24)& 18&0.5957(24)& 19&0.5387(36)\\
\hline \end{tabular} \end{center}}

Compare this with the inverse to the largest eigenvalue of our linear map 
$1/\Lambda=2/(3+\sqrt5)\approx0.38204$, which (see, for example,
\cite{BTV}) is the correct value of the ``second'' eigenvalue in this case. 

To show that even this is not the worst case we consider also another
partition, namely the standard partition shifted by $1/(2n)$ (in both 
directions). Similarly to the previous case, we associate the square 
centered at $(x,y)$ with the pair of integers $(i=nx, j=ny)$. Here 
$i,j \in \{0,1,\dots,n-1\}$. The transition probabilities for the 
corresponding Markov chain (whose elements are numbered as $jn+i+1$) are 
shown in the following table:

\begin{center}
\begin{tabular}{|c|c|c|c|c|c|} \hline
{}&$i+j,i+2j$&$i+j,i+2j+1$&$i+j+1,i+2j+1$&$i+j,i+2j-1$&$i+j-1,i+2j-1$\\ \hline
$i,j$& $1/2$    & $1/8$      & $1/8$        & $1/8$      & $1/8$  \\ \hline
\end{tabular} \end{center}

Moduli of the ``second'' eigenvalues ($r_2$) of the transition matrices 
($n^2 \times n^2$), and their multiplicities (in parenthesis) are 

\begin{center} \begin{tabular}{|r|l||r|l||r|l||r|l||r|l|} \hline
$n$& $r_2$& $n$& $r_2$& $n$& $r_2$& $n$& $r_2$& $n$& $r_2$ \\ \hline
 2 &0.3968(3 )& 3 &0.3953(8) & 4 &0.4543(12)& 5 &0.4029(20)& 6 &0.4443(24)\\ 
 7 &0.5577(16)& 8 &0.4940(24)& 9 &0.5038(24)& 10&0.4754(12)& 11&0.6203(20)\\
 12&0.4567(48)& 13&0.5371(28)& 14&0.5577(16)& 15&0.5495(16)& 16&0.4940(24)\\
 17&0.5864(36)& 18&0.6733(24)& 19&0.5976(36)& 20&0.4902(24)& 21&0.5838(32)\\ 
\hline \end{tabular} \end{center}

\?{\begin{center} \begin{tabular}{|r|l||r|l||r|l||r|l||r|l|} \hline
$n$& $r_2$& $n$& $r_2$& $n$& $r_2$& $n$& $r_2$& $n$& $r_2$ \\ \hline
 3 &0.3953(8) & 4 &0.4543(12)& 5 &0.4029(20)& 6 &0.4443(24)& 7 &0.5577(16)\\ 
 8 &0.4940(24)& 9 &0.5038(24)& 10&0.4754(12)& 12&0.4567(48)& 13&0.5371(28)\\
 14&0.5577(16)& 15&0.5495(16)& 16&0.4940(24)& 17&0.5864(36)& 18&0.6733(24)\\ 
\hline \end{tabular} \end{center}}

In this case all ``second'' eigenvalues are greater in modulus than 
$1/\Lambda$. Looking at these two tables the structure of limit points
of the eigenvalues seems not quite clear for both of the considered 
families of the partitions, and one might argue that for large enough
$n$ the corresponding $r_2$ may converge to $1/\Lambda$. However, we
have numerical evidence that for the standard partition for $n=7k$ 
there is an eigenvalue $0.4886$, while for the shifted standard
partition for $n=8k$ there is an eigenvalue $0.4940$. The following 
general statement justifies this prediction and provides us with the 
precise description of the structure of the spectrum for the case of 
a linear automorphism preserving integer points.

\begin{theorem} Let $\tilde f: \IR^d \to \IR^d$ be a linear map such
that $\tilde f(\IZ^d)=\IZ^d$, and let the map $f:=\tilde f \mod1$ be
defined on the $d$-dimensional unit torus. Denote by $P_n$ the matrix
corresponding to Ulam's approximation of the map $f$ constructed
according to the partition of the unit torus into $n^d$ equal cubes. 
Then $r \in \Sigma(P_{kn})$ for any positive integer $k$ whenever 
$r \in \Sigma(P_{n})$. \end{theorem}

The proof of this result is based on the fact that due to the
selfsimilar structure of Ulam's approximation in this case both
the matrix $P_{kn}$ and the eigenvector $e_{kn}$ corresponding to 
the eigenvalue $r$ consist of repeated blocks of the matrix $P_{n}$ 
and the eigenvector $e_{n}$ respectively.

Therefore if for some $n$ we obtain numerically a ``bad'' value for 
the ``second'' eigenvalue, it will still be present for large enough 
multiples of $n$.

In recent papers \cite{Fr2,BTV} it was proposed to use Ulam's
procedure based on a finite Markov partition to estimate $r_2$. 
This claim was justified in these papers for 
1D smooth expanding maps and 2D Anosov automorphisms. In practice,
the usefulness of this approach is limited by the observation that
usually such partitions can be found only 
numerically, and as we shall show a small error here may lead to even 
worse accuracy of the eigenvalues compared to a non Markov partition.

Now we are in a position to answer the question why the spectral gap
in the above numerics differs significantly from theoretical
predictions. To have a simple model for the analytical study to start
with, consider a family of 2D maps from the unit square into itself,
defined as follows: $f_\gamma(x,y):=(2x\mod1, \gamma (y-c)+c\mod1)$, 
$\gamma\ge0, \, 0<c<1$. For each value of $\gamma$ the map $f_\gamma$ is a
direct product of two 1D maps. Thus, for $\gamma>1$ one can prove that
the PF-spectrum of this map is just 
the set of all pairwise products of elements of the spectra of the 
involved 1D maps.
For $\gamma\le1$, however, only the first map
contributes to the spectrum. Observe that for $\gamma<1$ the invariant
measure is concentrated on the attracting fiber 
$\Gamma:=\{(x,c): \; 0\le x \le 1\}$, and the spectrum is the
PF-spectrum of the piecewise expanding map on $\Gamma$. The dependence
of $r_2$ on the parameter $\gamma$ is shown by thick lines in 
Figure~\ref{r2-direct}. Observe the discontinuous behaviour when 
the parameter $\gamma$ crosses the value $1$. By dots we
indicate the rate of correlations decay with respect to the Lebesgue
measure in this system. Observe that these two graphs differ only for
$0<\gamma<1$, i.e. when the stable foliation is present.

%%%%%%%%%%%%%%%%%%%%%%%%%%%%%%%%%%%%%%%%%%%%%%%%%%%%%%%%%%%%%%%%%%
%% $r_2(\gamma)$ for the direct product map.
\Bfig(160,110)
      {\bline(0,0)(1,0)(160) \bline(0,0)(0,1)(110)
       \put(150,0){\vector(1,0){10}} \put(0,100){\vector(0,1){10}}
       \bline(50,0)(0,1)(5) \bline(100,0)(0,1)(5) \bline(0,100)(1,0)(5)
       %\bline(0,50)(1,1)(50)
       \bezier{20}(0,50)(25,75)(50,100)
       \thicklines
       \bline(0,50)(1,0)(50) \bline(50,100)(1,-1)(50) 
       \bline(100,50)(1,0)(50)
       \put(2,-8){$0$} \put(48,-8){$1$} \put(98,-8){$2$} 
       \put(150,-8){$\gamma$}
       \put(-8,2){$0$} \put(-8,48){$\frac12$} \put(-8,98){$1$}
       \put(3,106){$r_2$}
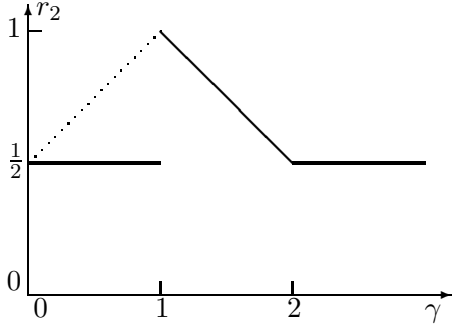
      }{$r_2(\gamma)$ for the direct product map. \label{r2-direct}}
%%%%%%%%%%%%%%%%%%%%%%%%%%%%%%%%%%%%%%%%%%%%%%%%%%%%%%%%%%%%%%%%%

This simple example demonstrates the main difference between 
expanding and hyperbolic maps, because the ``traditional'' spectrum 
in the latter case does not take into account the behaviour of the
system along the stable foliation.

\subsection{Random perturbations of contractive maps.}

The above examples show that in order to understand how small random 
perturbations change the behaviour of a
hyperbolic system one has to study their influence on a pure
contractive map. Let $f_{\gamma,c}(x):=\gamma(x-c)+c$, $0<\gamma,c<1$,
be a family of maps from the unit interval $[0,1]$ into itself. If our
random perturbation has the transition probability density
$q(\cdot,\cdot)$, being a \BV \ function of the first variable, then the
corresponding transition operator is well defined as an operator from
\BV \ into itself and one can compute its spectrum. Denote by $[x]$ the 
closest integer to the point $x$, and let $\delta=|[cn]-cn| \in
[0,1/2]$ be the distance from the fixed point $c$ to the closest
end-point of the Ulam interval to which it belongs, 
multiplied by $n$. The transition matrix
$P_n$ is lower triangular in this case. Therefore the diagonal
entries of the matrix are just its eigenvalues. A simple
calculation gives the following representation for $\tilde r_2$ as a 
function of $\delta$:
$$ \tilde r_2(\delta) = \function{
       2-\frac1\gamma
              &\mbox{if } \delta=0 \mbox{  and  } \gamma>\frac12  \\
       1-\delta(\frac1\gamma-1) 
              &\mbox{if } 0<\delta\le\frac{\gamma}{1-\gamma}\\
           0  &\mbox{otherwise} .} $$
This result shows the following. First, $\tilde r_2$ sensitively depends 
on the distance to the closest end-point of the Ulam interval it
       belongs to:
$\tilde r_2(0)=2-\frac 1\gamma$, 
$\tilde r_2(0_+)=1$, while $\tilde r_2(1/2)=(3-1/\gamma)/2$ 
(provided $\gamma>1/2$). 
Second, when the fixed point lies very close to the boundary of one of
the Ulam intervals, the estimate is the worst, which yelds a very bad
accuracy if one uses an approximation to a Markov partition for
Ulam's procedure.

Let us show that shift-invariant random perturbations
may cure this pathology. Suppose the random perturbation has a
shift-invariant transition probability density $q(x,y)=q(y-x)$, $q \in C^2$. 
We want to advocate that the PF-spectrum of the perturbed operator in zero 
noise limit in this case is well defined, does not depend on the shape of 
$q(\cdot)$ and is nontrivial. Let us prove this for $\tilde r_2$. 

\begin{lemma} Let $q \in C^2$, $q(x)=0$ if $|x|>\ep$, and 
$\ep\le\max\left\{\gamma c, \gamma(1-c)\right\}$. Then $\tilde r_2=\gamma$. 
\end{lemma}

\proof The random map can be rewritten as $x_{n+1}=\gamma (x_n-c) + c+ \xi_n$, 
where $(\xi_n)$ is a sequence of iid random variables with 
probability density $q(\cdot)$. Therefore 
$$ x_{n+1} = \gamma^n(x_1-c) + c 
           + (\gamma^{n-1}\xi_1 + \gamma^{n-2}\xi_2 + \dots + \xi_{n}) .$$
Let $\xi^{(n)}:=c+\sum_{k=0}^{n-1}\gamma^k\xi_{n-k}$.
Since the $\xi_k$ are iid, the sequence $(\xi^{(n)})$ converges in
distribution to a random variable $\xi^{(\infty)}$.
Then the random
variables $x_n$ converge in distribution to $\xi^{(\infty)}$ as
$n\to\infty$ and $\tilde r_2$ corresponds to the rate of this
convergence in the following sense: $\xi^{(\infty)}$ can be rewritten as
$\xi^{(\infty)} {\buildrel \rm d \over =} 
 \gamma^n (\tilde\xi^{(\infty)} -c) + \xi^{(n)}$,
where ${\buildrel \rm d \over =}$ means equality in distribution 
and $\tilde\xi^{(\infty)}$ is a copy of $\xi^{(\infty)}$
which is independent of the $\xi^{(n)}$. 
Denote by $q_n, q_\infty$, $h_n$, $\phi_n$ and $\psi_n$ the densities of the 
random variables $\xi^{(n)}, \xi^{(\infty)}$, $x_n$, 
$\gamma^n(x_1-c)$ and $\gamma^n(\xi^{(\infty)}-c)$ respectively. 
Then $h_{n+1}=q_n\star\phi_n$ and $q_\infty=q_n\star\psi_n$, and the 
supports of $\phi_n$ and $\psi_n$ are of order $\gamma^n$.
Observing that for any $h \in C^1$
$$ \int|h(x+\delta) - h(x)|\, dx = (|\delta| + o(\delta)) \var(h) , $$
we conclude that for large $n$
$$ \nl{q_n - q_\infty} = \int|q_n(x) - q_n(x-y)|\psi_n(y)\,dy\,dx 
                       \le O(\gamma^n)\var(q_n), $$
$$ \var(q_n - q_\infty) = \int|q_n'(x) - q_n'(x-y)\psi_n(y)|\,dy\,dx 
                       \le O(\gamma^n)\var(q_n'), $$
and similarly
$$ \nl{q_n - h_{n+1}} \le O(\gamma^n)\var(q_n), \qquad
   \var(q_n - h_{n+1}) \le O(\gamma^n)\var(q_n). $$
As $\xi^{(n)}=\xi_n+\gamma\xi^{(n-1)}$, there is some probability density 
$h$ such that $q_n=q\star h$. Therefore $\var(q_n)\le\var(q)$ and 
$\var(q_n')\le\var(q')$ so that
$$ \nv{h_n - q_\infty} 
   \le O(\gamma^n) \cdot (\var(q) + \var(q')) . $$
It is easily seen that one can choose the initial density $h_1$ such that 
this order of convergence is attained (e.g.\ take $h_1$ close to a 
$\delta$-function). This yields the claim of the lemma. \qed

\bigskip
\n{\em Acknowledgements}. M.B. gratefully acknowledges support by the 
Volkswagen-Stiftung and by INTAS-RFBR 95-0723 and RFFI grants. G.K.
was partially supported by the DFG under grant Ke 514/3-2.

%%%%%%%%%%%%%%%%%%%%%%%%%%%%%%%%%%%%%%%%%%%%%%%%%%%%%%%%%%%%%%%%%
\bigskip %\newpage


\begin{thebibliography}{99} \label{bibl}

\bibitem{Ba} V. Baladi, {\em Periodic orbits and dynamical spectra},
Preprint, 1997.

%{\em Dynamical zeta functions. Real and
%complex dynamical systems}, NATO Adv. Sci. Inst. Ser. C Math. Phys. 
%Sci., 464, Kluwer Acad. Publ., Dordrecht, 1995, 1--26. 

\bibitem{BY} V. Baladi, L.-S. Young, {\em On the spectra of randomly 
perturbed expanding maps}, Comm. Math. Phys. {\bf 156}:2 (1993),
355--385; {\bf 166}:1 (1994), 219--220.

\bibitem{Bl5} M.L. Blank, {\em Small perturbations of chaotic dynamical 
systems}, Uspekhi Matem. Nauk. {\bf 44}:6 (1989), 3--28. (English transl. 
Russ. Math. Surveys {\bf 44}:6 (1989), 1--33.)

\bibitem{Bl17} M.L. Blank, {\em Chaotic maps and stochastic Markov chains}, 
Abstracts of Congress IAMP--91, 1992, 6p.

\bibitem{Bl20} M.L. Blank, {\em Discreteness and continuity in problems of 
chaotic dynamics}, Monograph, Amer. Math. Soc., 1997.

\bibitem{BK95} M.L. Blank, G. Keller, {\em Stochastic stability versus 
localization in chaotic dynamical systems}, Nonlinearity {\bf 10}:1 (1997), 
81-107.

\bibitem{BTV} F. Brini, G. Turchetti, S. Vaienti, {\em Decay of correlations 
for the automorphism of the torus $T^2$}, Preprint Luminy, 1997.

\bibitem{dellnitz} M. Dellnitz, O. Junge, {\em On the approximation of 
complicated dynamical behavior}, to appear in SIAM Journal on Numerical 
Analysis, 1998.

\bibitem{Fr2} G. Froyland, {\em Computer-assisted bounds for the rate of 
decay of correlations}, Comm. Math. Phys. {\bf 189}:1 (1997), 237-257.

%\bibitem{Hu} F. Hunt, {\em A Monte Carlo approach to the approximation of 
%invariant measures}, Random \& Computational Dynamics 
%{\bf 2}:1 (1994), 111--133.

\bibitem{Ke1} G. Keller, {\em Stochastic stability in some chaotic dynamical
systems}, Mh. Math. {\bf 94} (1982), 313--333.

\bibitem{Ke2} G. Keller, {\em On the rate of convergence to equilibrium in 
one-dimensional systems}, Comm. Math. Phys. {\bf 96}:2 (1984), 181--193.

\bibitem{Ki3} Yu. Kifer, {\em Random perturbations of dynamical systems}, 
Boston: Birkhauser, 1988.

\bibitem{Ki4} Yu. Kifer, {\em Computations in dynamical systems via random 
perturbations}, Discrete Contin. Dynam. Systems {\bf 3}:4 (1997), 457--476.

%\bibitem{LY} A. Lasota, J.A. Yorke, {\em On the existence of invariant 
%measures for piecewise monotone transformations}, Trans. Amer. Math. Soc. 
%{\bf 186} (1973), 481--488.

\bibitem{Li}  T.Y. Li, {\em Finite approximation for the Frobenius-Perron 
operator. A solution to Ulam's conjecture}, J. Approx. Th. {\bf 17} (1976), 
177--186.

\bibitem{Ul} S. Ulam, {\em Problems in modern mathematics}, Interscience 
Publishers, New York, 1960.

\bibitem{Yo} L.-S. Young, {\em Statistical properties of dynamical systems 
with some hyperbolicity}, Preprint UCLA, 1996

\end{thebibliography}
\end{document}